%% file: prl_v3_final.tex
\documentclass[floatfix,twocolumn,prl,showpacs,preprintnumbers,amsmath,nofootinbib,amssymb,noeprint,superscriptaddress]{revtex4-1}
\usepackage[utf8]{inputenc}
\usepackage{comment}
\usepackage{bm}
\usepackage[pdftex]{color}
\usepackage{xcolor}
\usepackage{graphicx,capt-of}
\usepackage{amsmath,amssymb,amsfonts,mathtools}
\usepackage{braket}
\usepackage[colorlinks=true,linkcolor=black,filecolor=black,urlcolor=blue,citecolor=blue,pdftex,plainpages=false]{hyperref}

\usepackage{ulem}
\newcommand{\fourpt}{{4\mathrm{pt}}}
\newcommand{\twopt}{{2\mathrm{pt}}}
\newcommand{\Op}{\mathcal{O}}
\newcommand{\Opbar}{\overline{\Op}}
\newcommand{\tr}{\operatorname{tr}}
\newcommand{\dd}{\mathrm{d}}

\begin{document}
	\preprint{MIT-CTP/6027}
	
	\title{Parton Distribution Functions from Large Momentum Expansion \\ of Current-Current Correlators}
	
	\author{Jia-Lu Zhang}
	\affiliation{Tsung-Dao Lee Institute and School of Physics and Astronomy, Shanghai Jiao Tong University, Shanghai 201210, China}
	\author{Xiangdong Ji}
	\affiliation{Tsung-Dao Lee Institute and School of Physics and Astronomy, Shanghai Jiao Tong University, Shanghai 201210, China}
	\author{Andreas Sch\"afer}
	\affiliation{Institute for Theoretical Physics, University of Regensburg, 93040 Regensburg, Germany}
	\author{Rui Zhang}
	\email{rzhang93@mit.edu,corresponding author}
	\affiliation{Center for Theoretical Physics - a Leinweber Institute, Massachusetts Institute of Technology, Cambridge, MA 02139, USA}
	\author{Christian Zimmermann}
	\email{chrzim@lbl.gov,corresponding author}
	\affiliation{Department of Physics and Astronomy, University of Kentucky, Lexington, KY 40506, USA}
	\affiliation{Nuclear Science Division, Lawrence Berkeley National Laboratory, Berkeley, CA 94720, USA}
	\begin{abstract}
		The universality of the large momentum effective theory (LaMET) allows computing parton distribution functions starting from any Euclidean correlator with appropriate large momentum Fourier components. Nevertheless, only the Wilson-line operators have been widely studied in this framework.
		In this work, we provide 
		{the first}
		LaMET formulation using current-current correlators
		{up to the next-to-leading order}. The advantage 
		of this new method is its simple renormalization
		property due to the absence of Wilson-line linear divergences.
		This provides an alternative approach for LaMET calculation with very different systematics, making it crucial for cross-validating future precision calculations. A first numerical example is presented using newly generated correlators for specific LaMET application from existing four-point function data.
	\end{abstract}
	
	\maketitle
	\text{\it Introduction:}
	Parton distribution functions (PDFs) are fundamental quantities in quantum chromodynamics (QCD) that describe the hadronic structure in terms of quark and gluon degrees of freedom on the light cone. They are universal inputs to the prediction of hadronic processes at, e.g., the LHC~\cite{ATLAS:2019mfr} and the upcoming Electron-Ion Collider (EIC)~\cite{Accardi:2012qut,AbdulKhalek:2021gbh}, and can be determined from global fitting to experimental data.
	Meanwhile, it is also important to theoretically calculate these observables non-perturbatively from first-principles lattice QCD. While there are fundamental difficulties in fully extracting these light-like observables on a Euclidean lattice, large momentum effective theory (LaMET)~\cite{Ji:2013dva,Ji:2014gla} provides a framework to directly calculate the mid-range Bjorken-$x$ dependence of PDFs through the large momentum expansion of Euclidean correlations. Various other approaches have also been proposed to fit the parametrization of PDFs from lattice QCD data, which can be found in recent reviews~\cite{Lin:2017snn,Cichy:2018mum,Constantinou:2020pek,Constantinou:2022yye,Lin:2025hka}.
	
	The LaMET framework introduces a universality class of distributions~\cite{Ji:2020ect}. The leading order expansion of these distributions in terms of $(\Lambda_{\rm QCD}/P_z)$ contains the same infrared (IR) structure as the lightcone PDFs and thus can be perturbatively matched to them. The simplest member of this universality class is the quasi-PDF, which is the Euclidean version of the lightcone correlation function, and can be interpreted as the physical momentum distribution of quarks and gluons in a hadron with finite momentum. Other variants in the same universality class include the Coulomb gauge distributions proposed recently~\cite{Gao:2023lny,Zhao:2023ptv} and the current-current correlations~\cite{Ji:2020ect}. Since the proposal of LaMET, significant theoretical improvements have been developed to better control the systematics for the quasi-PDF calculations in the past decade~\cite{Ji:2017oey,Ishikawa:2017faj,Green:2017xeu,Constantinou:2017sej,Stewart:2017tvs,Alexandrou:2017huk,Izubuchi:2018srq,Li:2020xml,Chen:2020ody,Ji:2020brr,Gao:2021hxl,LatticePartonLPC:2021gpi,Su:2022fiu,Zhang:2023bxs,Ji:2023pba,Liu:2023onm,Ji:2024hit,Chen:2025cxr,Ji:2026vir}. One significant challenge of the precision PDF calculation from quasi-PDFs is the renormalization of the ultraviolet (UV) divergence in the Wilson-line self energy~\cite{Chen:2016fxx,Green:2017xeu,Ji:2017oey,Ishikawa:2017faj,Ji:2020brr}. Coulomb gauge distributions avoid this issue by not including the spatial Wilson line in the operator, and have demonstrated advantages in numerical precision in the calculation of the transverse-momentum-dependent  observables~\cite{Gao:2023lny,Zhao:2023ptv}, but their factorization has not been proved to all orders in perturbation theory. The current-current correlators have not yet been studied in LaMET applications due to the perception that they are more challenging to compute on lattice. 
	With continuing advances in computational resources, these correlators are becoming increasingly accessible in lattice QCD. 
	They offer the significant advantage of being free from renormalization issues and being well understood in perturbation theory.
	Consequently, they provide an alternative method with completely different systematics, which is substantially important for cross-validation on the control of systematic uncertainties in precision PDF.
	
	Current-current correlations have attracted a lot of interest due to their close relation to physical cross sections in collider physics. Various methods of extracting PDFs from current-current correlations %\XJ{using operator product expansion (OPE)} 
	have been proposed, including the hadronic tensor~\cite{Liu:1999ak}, the ``Ioffe time'' distributions~\cite{Braun:2007wv}, the Compton amplitude~\cite{Chambers:2017dov}, and the ``lattice cross sections''~\cite{Ma:2017pxb}. They can also be studied to understand the twist-4 double-parton distributions~\cite{Diehl:2011yj}. 
	Some first lattice calculations have been realized in recent years~\cite{Chambers:2017dov,Bali:2018spj,Liang:2019frk,Sufian:2020vzb,Bali:2021gel,Zimmermann:2024zde}. 
	The common basis of these methods for extracting PDFs is the operator product expansion (OPE), which allows a short distance factorization (SDF) of the correlators into lightcone distributions and perturbative matching kernels when the two currents are separated by a perturbative Euclidean distance. They can extract moments of PDFs, or fit the data to some parametrization of the $x$-dependent PDFs.
	
	As current-current correlations are also in the same universality class of distributions introduced by LaMET, lightcone PDFs can be factorized in terms of current-current correlations, and their $x$-dependence can be directly calculated through perturbative matching~\cite{Ji:2024oka}. However, here for the first time, we present the detailed formulation of this factorization up to one-loop order. The differences between SDF and LaMET applications of similar lattice data are explained in Ref.~\cite{Ji:2022ezo}.
	
	In this paper, we develop the $x$-space formulation for the large momentum expansion of the two-current matrix elements of vector (V) and axial vector (A) currents 
	{up to one-loop order for the first time, enabling future calculations in this direction. }
	For this first analysis, we generate new correlators specifically for the LaMET formulation from the current-current 
	data produced in 
	Refs.~\cite{Bali:2021gel}. We show some first results for the proton PDF $f_{u-d}$, demonstrating the feasibility of the presented approach.
	
	\text{\it Theory:} 
	We start from forward matrix elements of two currents in the proton. We consider unpolarized matrix elements, i.e.\ we implicitly average over helicity states:
	\begin{align}
		M_{XY,ab}^{\mu\nu}(z,P) 
		= 
		\bra{P}
		J^\nu_{X,ab}(z)\ 
		J^\mu_{Y,ba}(0)
		\ket{P} \,,
		\label{eq:Mdef}
	\end{align}
	where $J^\mu_{X,ab}(z)$ are local currents. Specifically, we consider two types: 	
	\begin{align}
		J^\mu_{\mathrm{V},ab}(z) 
		:=
		\bar{q}_a(z)\ \gamma^\mu\ q_b(z)
		\,,\nonumber\\
		J^\mu_{\mathrm{A},ab}(z) 
		:=
		\bar{q}_a(z)\ \gamma^\mu \gamma_5\ q_b(z) \,.
	\end{align}
	The indices $a$ and $b$ indicate the flavor of the quark field, with the index $a$ specifying the quark PDF under consideration. 
	We are free to select the flavor $b$, and the calculation simplifies drastically if we choose $b$ to be an (auxiliary) quark flavor that is different from the valence quark flavors of the considered hadron, in our case a proton. The spatial separation $z=(0,\vec{z})$ is chosen to be parallel to the direction of the proton's momentum $\vec{P}$ with $|\vec{P}|\gg\Lambda_{\rm QCD}$.
	In the following, we discuss the current combination VV where we already have lattice data from previous work, the case AA works in an analogous way. 
	The leading contribution to the hadronic tensor is given by transverse polarizations, thus for arbitrary lattice momentum $\vec{P}=(P_x,P_y,P_z)$, the spin structure can be replaced with the following projection of Dirac matrices,
	\begin{align}
		\Gamma^i=\sum_{j\in\{x,y,z\}}\left(\delta_{ij}-\frac{P_i P_j}{|\vec{P}|^2}\right)\gamma_j,
		\label{eq:transverse_proj}
	\end{align}
	where only two of them are independent, spanning the 2-dimensional transverse plane, such that $\Gamma\cdot \vec{P}=0$. We can further average the data on the transverse plane for the VV current by taking a trace, yielding
	\begin{align}
		M^{\perp\perp}_{\mathrm{VV},ab}&(z,P)
		=\frac{1}{2}\sum_{i,j} \left(\delta_{ij}-\frac{P_iP_j}{|P|^2}\right)
		M_{\mathrm{VV},ab}^{ij}(z,P)
		\label{eq:Wperp}
	\end{align}
	Moreover, it is useful to define:
	\begin{align}
		W^{\perp\perp}_{\mathrm{VV},ab}(z,P) := 
		-\frac{i\pi^2 z^4}{\vec{z}\cdot\vec{P}}
		M^{\perp\perp}_{\mathrm{VV},ab}(z,P) \,,
		\label{eq:Wdef}
	\end{align}
	and likewise for the AA combination.   As explained in Ref.~\cite{Ma:2014jla}, these correlators exhibit the same infrared divergence structure as the PDFs.
	Therefore, their $x$-space conjugate can be used to calculate the PDFs through the LaMET approach. 
	One of the most important merits of these correlators is that they are constructed from conserved vector and axial-vector currents and do not involve Wilson lines. As a result, no UV divergence arises at one-loop order, and the correlators are free from power UV divergences and the linear renormalon ambiguities.
	
	As we shall explain later in this work, we average the matrix elements of the VV and AA currents,
	\begin{align}
		W^{\perp\perp}(z,P)=\frac{1}{2}\left(	W^{\perp\perp}_{\mathrm{VV}}(z,P)+	W^{\perp\perp}_{\mathrm{AA}}(z,P)\right),
		\label{eq:VVAA-combo}
	\end{align}
	and Fourier transform to obtain the VV-distribution,
	\begin{align}
		\tilde{h}(y,P_z)
		&=
		\int_{-\infty}^\infty \frac{P_z\dd z}{2\pi}\,
		e^{i y zP_z}\,
		W^{\perp\perp}\!\left(z,P\right),
	\end{align}
	which can also be directly calculated in momentum space ~\cite{Grebe:2026qmt}.
	The light-cone PDF can be factorized into the VV-distribution and perturbative matching~\cite{Ji:2024oka},
	\begin{align}
		\label{eq:lm_expansion}
		f(x,\mu^2)
		=
		\int_{-\infty}^{\infty} \frac{\dd y}{|y|}\,
		C\!\left(\frac{x}{y},\frac{|y|P_z}{\mu}\right)
		\, \tilde{h}(y,P_z)
		+ ...\, ,
	\end{align}
	where $C\!\left(\frac{x}{y},\frac{|y|P_z}{\mu}\right)$ is the matching kernel calculable in
	perturbation theory, and the higher twist contributions in the expansion parameter $\mathcal{O}\left(\frac{\Lambda^n_{\rm QCD}}{(xP_z)^n},\frac{\Lambda^n_{\rm QCD}}{((1-x)P_z)^n},\frac{M^n}{P_z^n}\right)$ are omitted for $n\geq2$. In this way, LaMET provides a forward matching
	that enables a direct determination of PDFs from lattice-computable
	quasi-distributions, with systematically improvable power corrections suppressed by inverse powers of $P_z$.
	
	We calculate the perturbative matching kernel up to one loop in the $\overline{\text{MS}}$ scheme, with the \text{\it naive} \text{\it dimensional regularization}, where the VV and AA results are exactly  the same at one loop order. Thus we have,
	\begin{widetext}
			\begin{align}\label{eq:matching}
				C_{\rm ns}^{\overline{\text{MS}}}\bigg(\xi,\frac{|y|P_z}{\mu}\bigg)=&\delta(1-\xi)\left(1+\frac{\alpha_s C_F}{4\pi}\right) + \frac{\alpha_s C_F}{2\pi} \left\{\frac{\left(\xi ^2+1\right)[{\rm sgn(1-\xi)}{\rm sgn(\xi)} \log |\xi| + \log |1-\xi|]+\xi +1/2 }{-|\xi-1|}\right\}_{+(1)}^{[-\infty,\infty)}\nonumber\\
				&+\frac{\alpha_s C_F}{2\pi}
				\theta(\xi)\theta(1-\xi)   \left\{\frac{1+\xi^2}{1-\xi}\log\bigg(\frac{\mu^2}{4y^2P_z^2}\bigg)+ \bigg[\frac{2\log (1-\xi )+ (\xi -5) \xi +3}{ (\xi -1)}\bigg]\right\}_{+(1)}^{[0,1]}
        \end{align}
	\end{widetext}
	where $\xi=\frac{x}{y}$, and the plus distribution is defined as
	\begin{align}
		\int \dd x [g(x)]^{D}_{+(x_0)} h(x)
		=
		\int_D \dd x g(x)[h(x)-h(x_0)].
	\end{align}

    The PDFs can also be calculated from the VA or AV correlations, which have been studied before for the pion valence PDF in the short-distance factorization framework~\cite{Sufian:2020vzb}. We can obtain the VA matrix elements through (assuming $x,y$ to be two transverse directions),
	\begin{figure}[tb]
		\centering
		\includegraphics[width=\columnwidth]{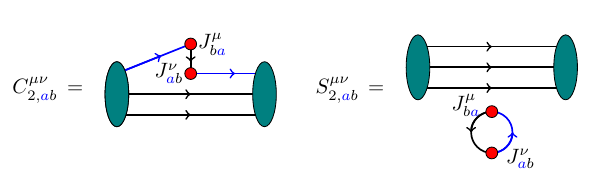}
		\caption{Wick contractions that can contribute to the four-point correlation function \eqref{eq:4ptdef} in the case of an auxiliary intermediate quark field of flavor $b$. Flavor $a$ (blue) specifies the PDF.}
		\label{fig:graphs}
	\end{figure}
	\begin{align}
		W^{\perp\perp}_{\mathrm{VA}}:=\frac{\pi^2z^3}{2E_P}
		&\left[
		M^{xy}_{\mathrm{VA}}(z,P)
		+
		M^{xy}_{\mathrm{AV}}(z,P)
		\right]\,,
	\end{align}
	where $E_P = \sqrt{\vec{P}^2 + m^2}$ is the proton energy.
	The corresponding matching kernel needs an extra correction term compared to Eq.~\eqref{eq:matching},
	\begin{align}
		C_{\rm VA, ns}^{\overline{\text{MS}}}=C_{\rm ns}^{\overline{\text{MS}}}+\frac{\alpha_sC_F}{\pi}[1-\xi]_{+(1)}^{[0,1]}\theta(\xi)\theta(1-\xi)\,.
	\end{align}
	
	\text{\it Simulation details:}
	In this first analysis, we generate the matrix elements for the specific LaMET formulation from  the two-current data produced in the context of other simulations \cite{Bali:2021gel} employing the $n_f=2+1$ Wilson-Clover gauge ensembles provided by the CLS collaboration \cite{Bruno:2014jqa}. Specifically, we use the ensemble H102 with extension $32^3\times 96$, lattice spacing $a=0.085~\mathrm{fm}$, and pion mass $m_\pi=355~\mathrm{MeV}$. The matrix element in \eqref{eq:Mdef} can be calculated on the lattice in Euclidean spacetime if the separation between the two currents is purely spatial. The  matrix elements can then be expressed in terms of a Euclidean four-point correlation function in the limit of large time separations to suppress excited state contributions:	
	\begin{align}
		M^{\mu\nu}(z,P)
		=
		E_P V
		\left.
		\frac{C^{\mu\nu}_{\fourpt}(\vec{P},\vec{z},\tau,t)}{C_{\twopt}(\vec{P},t)}
		\right|_{0\ll\tau\ll t} \,,
		\label{eq:M-ratio}
	\end{align}
	where the four-point function reads
	\begin{align}
		&C^{\mu\nu}_{\fourpt}(\vec{P},\vec{z},\tau,t)
		\nonumber\\
		&\quad:=
		\left\langle
		\Op^{\vec{P}}(t)\ 
		J_{ab}^\mu(\vec{z},\tau)\ 
		J_{ba}^\nu(\vec{0},\tau)\ 
		\Opbar^{\vec{P}}(0)
		\right\rangle \,.
		\label{eq:4ptdef}
	\end{align}
	$C_{\twopt}$ is the usual two-point function, and $V$ is the spatial lattice volume. The operators $\Op^{\vec{P}}$, $\Opbar^{\vec{P}}$ respectively annihilate or generate the desired hadron, in our case, the proton. In general, there are several types of Wick contractions that can contribute to \eqref{eq:4ptdef} as has been worked out in detail in \cite{Bali:2021gel}. As already mentioned, we consider $b$ to be different from the flavors of the valence quarks in the considered hadron (non-singlet contribution). In this case, only the contractions $C_2$ and $S_2$ shown in Fig. \ref{fig:graphs} contribute. The latter represents a pure sea quark contribution. In the current analysis, we restrict ourselves to the isovector PDF, i.e.\ $f_{u-d}$, where the $S_2$ contributions cancel. Hence, we only have to evaluate the contraction $C_2$. The methods and techniques to evaluate the required diagrams have been worked out in \cite{Bali:2021gel}.

	In this work, we only give a brief summary of the required diagram $C_2$: We use a momentum smeared source \cite{Bali:2016lva} to obtain a point-to-all propagator. The sink is realized using the sequential source technique again employing momentum smearing.
	The propagator between the two currents is evaluated using the stochastic source technique,  accompanied by  the hopping parameter expansion for Wilson fermions. %\CZ
	{For more details on this, the reader is referred to \cite{Bali:2009hu,Bali:2021gel}}. This allows us to efficiently calculate the matrix element \eqref{eq:M-ratio} for all current separations $\vec{z}$. 
	As discussed in \cite{Zimmermann:2024zde}, the contribution of the $C_2$ diagram is sensitive to discretization artifacts of the Wilson propagator {partly} due to a non-vanishing chiral odd contribution to the Wilson propagator. This is avoided by taking combinations of Dirac structures where these chiral odd contributions cancel, e.g.\ the combination $(M_{VV}+M_{AA})/2$ given in \eqref{eq:VVAA-combo} \cite{Bali:2018spj,Zimmermann:2024zde}. The remaining anisotropy artifacts are suppressed by performing a simple tree-level improvement where we multiply our data by the correction factor:
	\begin{align}
		c^{\mathrm{corr}}(z) 
		=
		\frac{\tr\left\{ z\!\!\!/ M^{\mathrm{free}}_{\mathrm{cont}}(z) \right\}}{\tr\left\{ z\!\!\!/ M^{\mathrm{free}}_{\mathrm{latt}}(z) \right\}}
		= 
		-\frac{m_q^2}{\pi^2} \frac{K_2(m_q\sqrt{-z^2}) }{ \tr\left\{ z\!\!\!/ M^{\mathrm{free}}_{\mathrm{latt}}(z) \right\} } \,,
	\end{align}
	{where $m_q = \frac{1}{2}(m_{u} +m_{d})m_{\pi,\mathrm{H102}}^2/m^2_{\pi_0} \approx 24~\mathrm{MeV}$ is the $\overline{\rm MS}$ quark mass in our lattice setup for the above equation. }
	In order to keep the {associated} systematic uncertainties  small, it is advisable to restrict to data points where $c^{\mathrm{corr}}(z)$ is close to 1. In past studies, it has been found  {to be} the case if $\vec{z}$ is in the vicinity of  {a lattice diagonal}  \cite{Bali:2018spj,Zimmermann:2024zde}. The simulation is carried out for proton momenta of the form $(\pm n,\pm n,\pm n)$ so that $\vec{z}$ is indeed aligned with one of the lattice diagonals if one maximizes the scalar product $\vec{P}\cdot \vec{z}$, which is preferred in our LaMET calculation. At the moment, the largest available proton momentum is $|\vec{P}|_{\mathrm{max}} = \sqrt{12}\times 2\pi/L = 1.57~\mathrm{GeV}$.

\begin{figure}
		\centering
		\includegraphics[width=0.9\columnwidth]{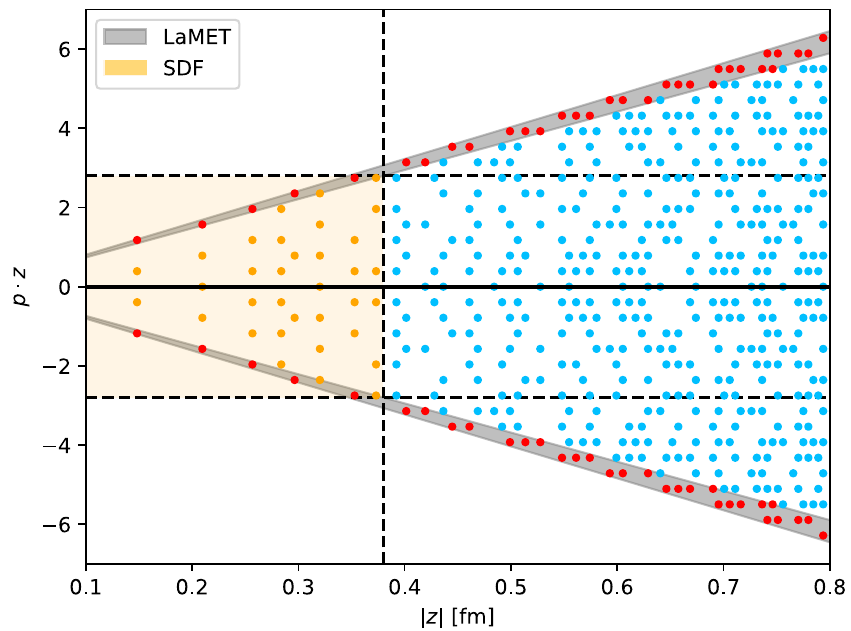}
		\caption{Kinematic regions of the four-point function data used in the LaMET and the SDF analysis.}
		\label{fig:kinematical_region}
	\end{figure}

	To maximally utilize the lattice data we have measured, we also include more data points when $\vec{P}$ and $\vec{z}$ are not perfectly aligned, and also other relevant Lorentz components in addition to the pure transverse operators. This increases the effective statistics of the measurement, and allows a smoother interpolation in coordinate space and a more stable Fourier transformation to $x$-space. To extract $M^{\perp\perp}$, we exploit the fact that $M^{\mu\nu}$ is a Lorentz-covariant object and decompose it in the case of $M_{VV}$ and $M_{AA}$ into the Lorentz structure:	
	\begin{align}
		&M^{\mu\nu}(z,P)
		=
		\frac{P^\mu P^\nu}{m^2} A(\lambda,z^2)
		-
		g^{\mu\nu} B(\lambda,z^2)
		\nonumber\\
		&\quad+
		m^2 z^\mu z^\nu C(\lambda,z^2)
		+
		(P^\mu z^\nu + P^\nu z^\mu) D(\lambda,z^2)\,,
		\label{eq:Ldecomp}
	\end{align}
	%
	%\CZ
	{where $\lambda = \vec{P}\cdot\vec{z}$.}
	$M^{\perp\perp}$ is equivalent to the function $B$, {which is {obtained} from the over-determined system of Eq.~\eqref{eq:Ldecomp} including all available data points.}
	$B$ is accessible for $-|\vec{P}|_{\mathrm{max}}|\vec{z}| \le \lambda \le |\vec{P}|_{\mathrm{max}}|\vec{z}|$. 
	By definition, the actual $P_z$ is now $P_z=\lambda/|\vec{z}|$ when $\vec{z}$ is not aligned with $\vec{P}$. To avoid a large logarithmic correction from combining data at different $P_z$, we only allow a small tolerance of $2\%$ with respect to \ $|\vec{P}|$, i.e.\ $|\vec{P}|(1-0.02) \le \lambda/|\vec{z}| \le |\vec{P}|(1+0.02)$.
	{Notice that the corresponding dataset is essentially different from the one required for SDF approaches. The latter require data points for all $\lambda$ at fixed small $z^2$, while in the current work we need the information for maximal $\lambda$ for all available $z^2${, as demonstrated in Fig.~\ref{fig:kinematical_region}.}} 
	
\begin{figure}[tb]
		\centering
		\includegraphics[width=\columnwidth]{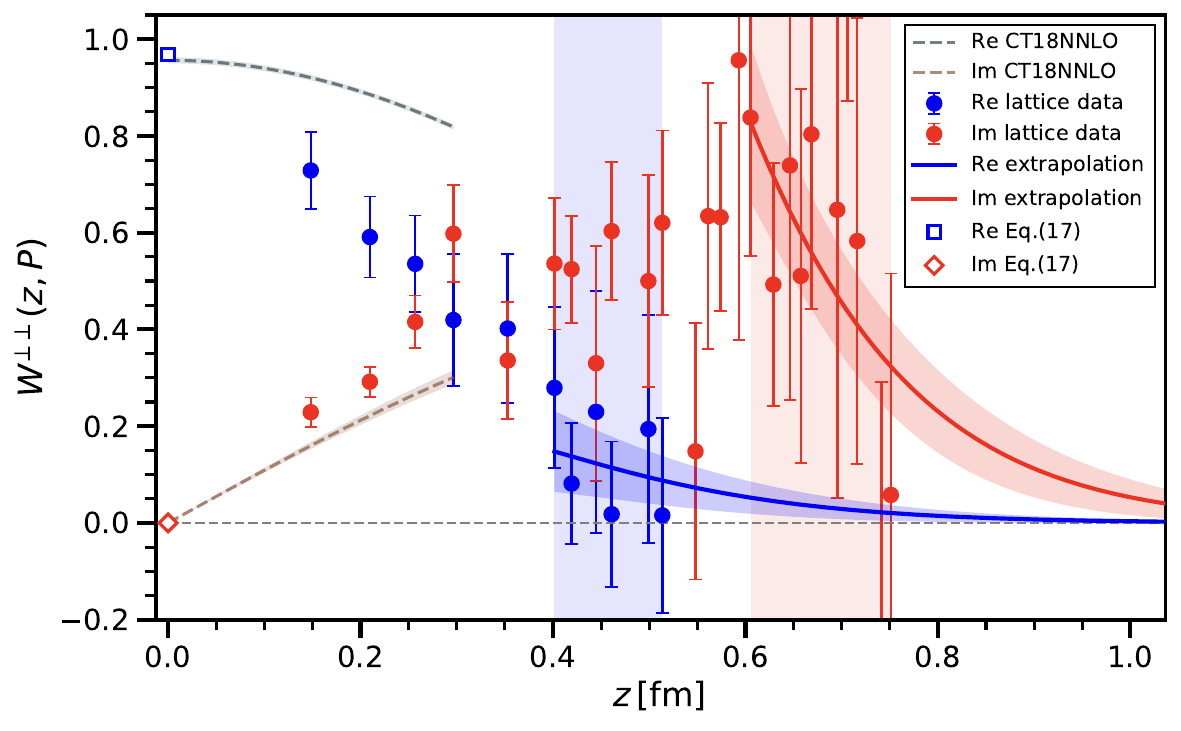}
		\caption{Real (blue) and imaginary (red) parts of $W_{u-d}^{\perp\perp}(z,P)$ at $|\vec P|=1.57~\mathrm{GeV}$. 
			{The vertical shaded bands label the extrapolation regions, and the dashed curves are reconstructed from CT18NNLO through {coordinate-space} matching in the perturbative region.}
		}
		\label{fig:raw_data}
	\end{figure}

	\text{\it Lattice Results:}
	The lattice results for $W^{\perp\perp}_{u-d}$ are shown in 
	Fig.~\ref{fig:raw_data} as functions of $z$. Given the current level of statistics, the real part of the matrix elements becomes consistent with zero at around $z\approx 0.42$ fm. %\ZJL
	{In addition, we compare the lattice data with $W_{u-d}^{\perp\perp}(z,P)$ reconstructed from the CT18 PDF~\cite{Hou:2019efy} through  {coordinate-space} matching. The CT18NNLO result is shown only in the perturbative region $z\leq0.3~\mathrm{fm}$ where the SDF is reliable. 
		The deviation suggests larger moments of the PDF in our lattice data.
	} For the moment, we employ a suitable ansatz for extrapolating the large-$z$ behavior. This generates some systematic uncertainty, which should be suppressed by improved data quality at large $z$ in future simulations.
	The matrix element at $z=0$ is not directly accessible in lattice 
	simulations because of the contact singularity. However, since the 
	current-current operator is free of ultraviolet divergences, the 
	corresponding $z=0$ value can be calculated perturbatively. In the 
	$\overline{\mathrm{MS}}$ scheme, one finds 
	\begin{align}
		W^{\perp\perp}(0,P)
		=
		1-{\alpha_sC_F}/{4\pi},
		\label{eq:W0def}
	\end{align}
	which can then be included in the Fourier transform.
	
\begin{figure}[tbp]
		\centering
		\includegraphics[width=\columnwidth]{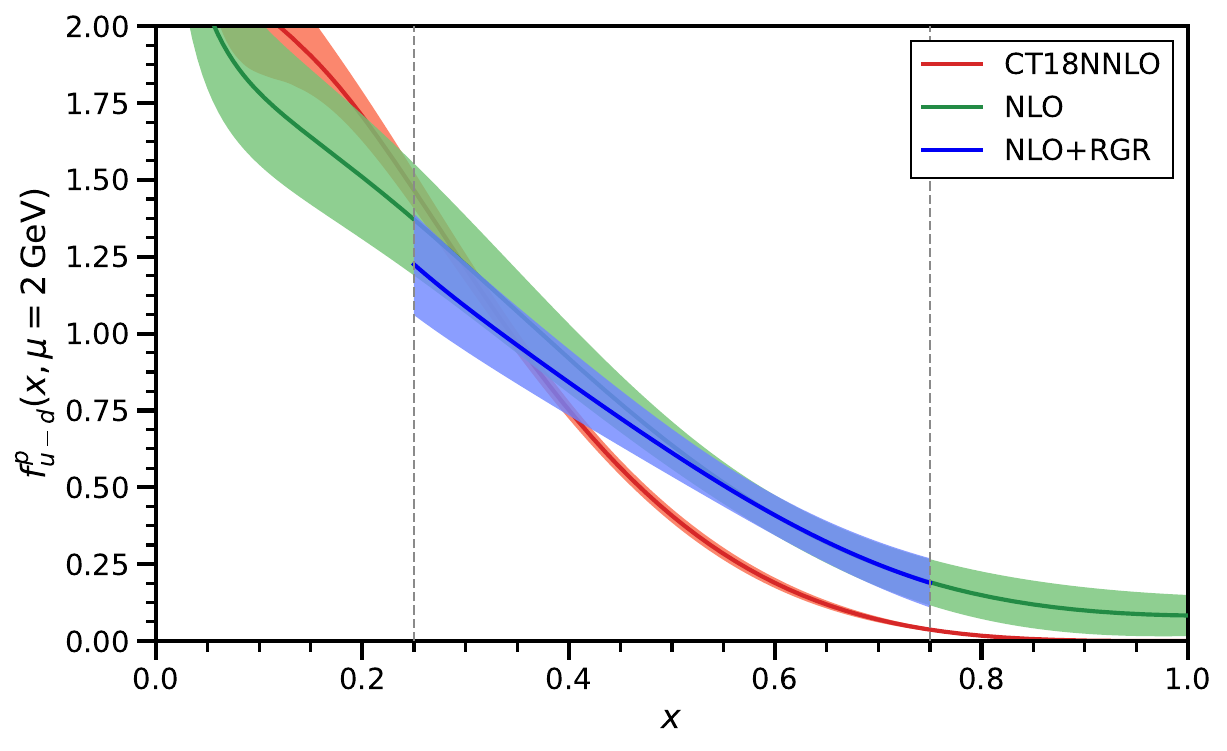}
		\caption{The flavor non-singlet PDF $f_{u-d}(x,\mu)$ at 
			$\mu=2~\mathrm{GeV}$. The blue band shows the result obtained in 
			this work, while the red band denotes the CT18NNLO global-fit 
			result.}
		\label{fig:compare}
	\end{figure}

	To carry out the Fourier transform, we adopt an extrapolation ansatz 
	for the large-$|z|$ region, motivated by its asymptotic behavior derived in Ref.~\cite{Ji:2026vir}. 
	Specifically, we parameterize the correlator as
	\begin{equation}
		W^{\perp\perp}(z,P)=A\,|z|^{5/2}e^{i\phi\,\mathrm{sign}(z)}e^{-\Lambda|z|},
	\end{equation}
	where $A$, $\phi$ and $\Lambda$ are fitting parameters.
	The extrapolation ansatz is used in the Fourier transform starting from the point where the lattice data become dominated by noise. This happens after $z=0.51~\mathrm{fm}$ for the real part and $z=0.75~\mathrm{fm}$ for the imaginary part. With this choice, the dominant uncertainty remains statistical rather than the model dependence of the extrapolation.
	Using this extrapolation together with the perturbative input at $z=0$, 
	we obtain the momentum-space distribution $\tilde{h}(y,P_z)$. 
	After applying the matching formula in Eq.~(\ref{eq:lm_expansion}), and also following the renormalization group resummation (RGR) procedure proposed in Ref.~\cite{Su:2022fiu},  the 
	resulting isovector PDF is shown in Fig.~\ref{fig:compare}, 
	together with the CT18NNLO global-fit result. A noticeable deviation 
	is observed. This is expected to be improved with larger hadron momenta, at which the parton's momentum will shift to smaller $x$. We emphasize that the present analysis with limited precision and hadron momentum is not intended to provide an 
	accurate determination. Rather, it aims to demonstrate a feasible method 
	for extracting PDFs from lattice QCD. 
	
	\text{\it Conclusion:} We formulated a large momentum expansion of two-current matrix elements of two vector or two axial vector currents in order to calculate the $x$-dependence of the proton PDF. 
	This first analysis was carried out for $f_{u-d}$, avoiding disconnected diagrams. This work provides a proof of concept of using current-current correlations in the framework of LaMET, which has several conceptual advantages compared to the conventional approach using Wilson-line operators, such as the absence of the linear renormalon ambiguities. 
	%\CZ
	Therefore, this opens the door to a valuable alternative within the scope of LaMET with different systematics compared to the conventional Wilson-line approaches.
	Notice that our analysis was carried out at a maximal proton momentum $|\vec{P}|=1.57~\mathrm{GeV}$ due to limitations in the availability of lattice data, which we calculated from the raw data produced in the context of a previous project that was not designed for performing a LaMET calculation. 
	Moreover, the pion mass $m_\pi = 355~\mathrm{MeV}$ of the employed ensemble is higher than the physical one. This is in all probability the origin of the discrepancies between our lattice results and the experimental ones. In fact, as the hadron momentum increases, the partons are expected to migrate from large to small $x$ regions.
	%\CZ
	{Our formalism can be extended to quark PDFs of specific flavor by adapting the matching kernel accordingly. This would require the evaluation of the contraction $S_2$ in Fig.~\ref{fig:graphs}, which is non-trivial to implement, since it requires two all-to-all propagators.}
	The next step will be to repeat the calculations for higher momenta ($|\vec{P}| > 2~\mathrm{GeV}$) for the same ensemble. 
	In this context, the usage of the newly proposed kinematically enhanced interpolators \cite{Zhang:2025hyo} will be very helpful.
	Eventually we move on to other ensembles in order to take the physical and continuum limits.
	{Notice that the analysis can also be carried out for other hadrons, in particular the pion, where the signal-to-noise ratio is in general better~\cite{Bali:2018nde} {and gets a larger kinematical enhancement~\cite{Zhang:2025hyo}.}}

	\section*{Acknowledgments}
We thankfully acknowledge the CLS effort for generating the $n_f = 2 + 1$ ensembles in \cite{Bruno:2014jqa}, one of which was used for this work. X.-J. is partially supported by NSFC Grant No. 12635004, the State Key Laboratory of Dark Matter Physics and Thomas and Linda Lau Family Foundation. 
J.~Z. is supported by the T.~D.~Lee Scholarship. R.~Z. is partially supported by the U.S. Department of Energy, Office of Science, Office of Nuclear Physics under Contract No.~DE-SC0011090. R.~Z.  and C.~Z. are partially supported by DOE Quark-Gluon Tomography (QGT) Topical Collaboration under award No.~DE-SC0023646.  C.~Z. is supported in part by the Alexander von Humboldt Foundation and the U.S. Department of Energy, Office of Science, Office of Nuclear Physics, under Grant No.~DE-SC0013065 and under Contract No.~DE-AC02-05CH11231 which is used to operate Lawrence Berkeley National Laboratory.

%\bibliography{refs}
\input{prl_v3_final.bbl}
\end{document}

%% file: prl_v3_final.bbl
%Control: production of eprint (-1) disabled
%